# Design of CMOS-memristor Circuits for LSTM architecture

Kamilya Smagulova, Kazybek Adam, Olga Krestinskaya and Alex Pappachen James

Nazarbayev University
Astana, Kazakhstan, apj@ieee.org

**Abstract**

Long Short-Term memory (LSTM) architecture is a well-known approach for building recurrent neural networks (RNN) useful in sequential processing of data in application to natural language processing. The near-sensor hardware implementation of LSTM is challenged due to large parallelism and complexity. We propose a 0.18 µm CMOS, GST memristor LSTM hardware architecture for near-sensor processing. The proposed system is validated in a forecasting problem based on Keras model.

## Introduction

There are numerous applications that applies sequentially ordered data for prediction and classification using recurrent neural network (RNN). Feedback connections in RNN enables its units to retain previous information to compute the current stat, making RNN an efficient tool for processing sequential data where maintaining order of information is important. RNN training is performed using 'backpropagation through time' (BPTT) leading to either vanishing or exploding gradient problem. Long short-term memory (LSTM) (see Fig.1 is a extension of RNN that overcomes vanishing gradient problem [1,2].

## LSTM architecture

In Fig.1, the inputs and the outputs to the LSTM cell are vectors. Input $x(t)$ at time $t$ is of size $N$ and the rest of inputs and outputs are of size $M$ referring to the number of hidden LSTM units. The $m'$ refers to a single element in the vectors. For instance, $f_{m_t'}$ is the output of the forget gate for $m$'th element. $M$ elements form a vector [3]. The parameter $n$ is an input iterator index ($n \in [1,N]$), $m'$ is a current hidden unit index ($m' \in [1,M]$), and $m$ is a hidden unit iterator index ($m \in [1,M]$).

### A. Implementation of LSTM network for prediction problem

In [4], LSTM network is used to predict the number of international airline passengers. The network model such as for the regression problem is shown in the Fig.2 that consists of an input, a hidden layer comprised of a single LSTM cell with 4 units and an output layer with no activation function layer.

A single input value $x_i$ is applied to LSTM cell and its input gate forms the current intermediary cell state $c_t$. Initially, due to absence of a previous cell state data $C_{t-1}$ and previous cell output vector $h_{t-1}$, the present cell state $C_t$ becomes equal to $c_t$. The resulting cell state vector contribute to values of the final output vector of the LSTM unit $h_t$. The obtained vector $h_t$ passes through an output layer which produces a prediction of the next value.

### B. LSTM CMOS-memristive hardware implementation

The LSTM cell architecture is designed using the memristive crossbar and CMOS circuits. The proposed architecture computes the gate output values and the intermediary cell state $c_t$ per LSTM hidden unit $m'$ one step at a time. After $M$ cycles we obtain the full output vector for a LSTM cell. At each cycle obtained values are saved in the memory units.

Memristive crossbar implementation for matrix-vector multiplication [5] of a single LSTM cell with $N$ inputs and $M$ LSTM hidden units per cell is shown in Fig. 3. The inputs $x_i$, where $i \in [1,N]$, belong to the time step $t$, and the inputs $h_j$, where $j \in [1,M]$, are the outputs of corresponding LSTM blocks at time $t-1$. The biases for input, forget, and output
gates are shown as $b_i$, $b_f$, and $b_o$, respectively. The bias $b_c$ is for the intermediary cell state $c_t$. The real cell state is $C_t$. In Fig. 3, the weights of the inputs and biases are represented by the conductances of the memristors in the crossbar. Four structures delimited by the dashed blue lines compute the outputs of the gates: $i_t$, $f_t$ and $o_t$, and the intermediary cell state $c_t$. The transistors in the structures serve as switches that allow to perform weighted summation from a single crossbar column at a particular time step.

The resulting currents from the read transistors, that represent dot product of inputs and crossbar weights, are fetched into current mirror circuits. The mirrored currents are then fed to corresponding activation function circuits: sigmoid and hyperbolic tangent circuits. The activation function circuits produce the output voltage values for $i_t$, $f_t$, $c_t$ and $o_t$ required to obtain the cell state $C_t$. The cell step $C_{t-1}$ from the previous time step, is stored in a memory unit. Finally, the output $h_t$ of a current LSTM hidden unit is obtained as a voltage after few steps of conversions and calculations, performed in multiplier circuits, analog adder and voltage to current converter. The calculation of the final predicted output is obtained from the memory unit combining all hidden layer outputs $V_{h_M}$ and output layer weights to obtain dot product, which is represented as a current going through the resistor $R$ proportional to $V_{out}$.

## Results and discussion

The input dataset includes 144 observations during 12 years, was normalized to the range from 0 to 1 and divided into training and testing sets. Upon training of the network for 100 epochs, we extracted values of the network weight matrices for a constrained-range between -1 and 1. LSTM unit is composed of three matrices of sizes [1,16] for input signal $x_i$, [4,16] for recurrent connections with $h_*$ and biases $b_*$ [1,16]. Outputs of the columns in the matrix form input gate, forget gate, intermediary cell state and output gate vectors. The output layer has two matrices, e.g. weight matrix of size [4,1] and single base value 1. Obtained score for training set is smaller than for testing set (24.84% root-mean-square error (RMSE) against 55.26% RMSE) and requires further optimization.

To build a circuit, TSMC 0.18 µm CMOS technology and the memristor model for large scale simulations [6] was utilized. We use GST memristors that can be adjusted to 16 different resistance levels from $200k\Omega$ to $2000k\Omega$ [7], thus extracted weights interpolation was performed. Adjusting weights according to memristor conductance values also affects the network performance in either way - degrading or improving the prediction. Simulation with modified weights led to a slight improvement in performance (24.61% RMSE for training and 47.33% RMSE for testing sets). The total area and power consumption of a single LSTM unit is $77.326 \mu m^2$ and $105.9 mW$, respectively.

## Conclusion

In this work, the hardware architecture of a LSTM Keras model for forecasting was presented. Memristor crossbar array has high efficiency which allows to compute larger complex vector-matrix multiplication in single constant time step. However further network optimization and development of learning circuit is required.

## References

[1] K. Smagulova, O. Krestinskaya, and A.P. James. "A memristor-based long short term memory circuit." *Analog Integrated Circuits and Signal Processing* (2018): 1-6.
[2] F. A. Gers, J. Schmidhuber, and F. Cummins, "Learning to forget: Continual prediction with lstm," 1999.
[3] I. Goodfellow, Y. Bengio, A. Courville, and Y. Bengio, *Deep learning*. MIT press Cambridge, 2016, vol. 1.
[4] J. Brownlee, "Time series prediction with lstm recurrent neural networks in python with keras," Available at: *machinelearning-mastery. com*, 2016.
[5] M. Hu, J. P. Strachan, Z. Li, E. M. Grafals, N. Davila, C. Graves, S. Lam, N. Ge, J. J. Yang, and R. S. Williams, "Dot-product engine for neuromorphic computing: programming 1t1m crossbar to accelerate matrix-vector multiplication," in *Proceedings of the 53rd annual design automation conference. ACM*, 2016, p. 19.
[6] D. Biolek, Z. Kolka, V. Biolkova, and Z. Biolek, "Memristor models for spice simulation of extremely large memristive networks," in *2016 IEEE International Symposium on Circuits and Systems (ISCAS)*, May 2016, pp. 389–392.
[7] S. Xiao, X. Xie, S. Wen, Z. Zeng, T. Huang, and J. Jiang, "Gst-memristor-based online learning neural networks," *Neurocomputing*, 2017.

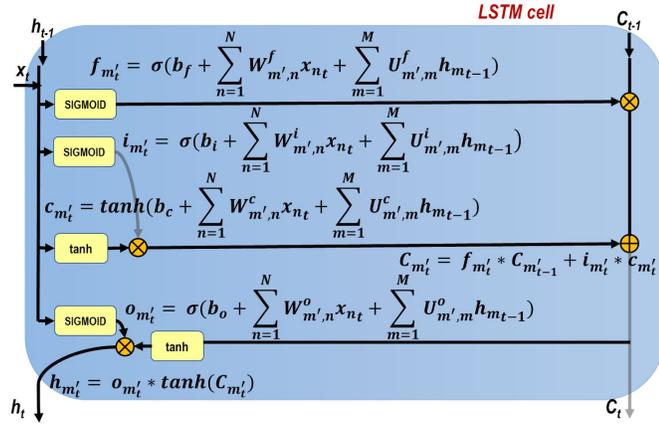

Fig. 1 Mathematical representation of conventional LSTM cell [2].

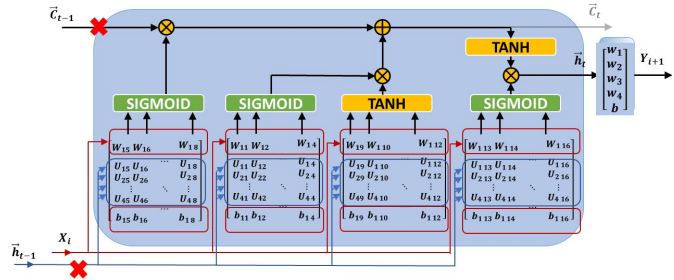

Fig. 2 Implementation of LSTM network for the particular example of prediction making problem [3].

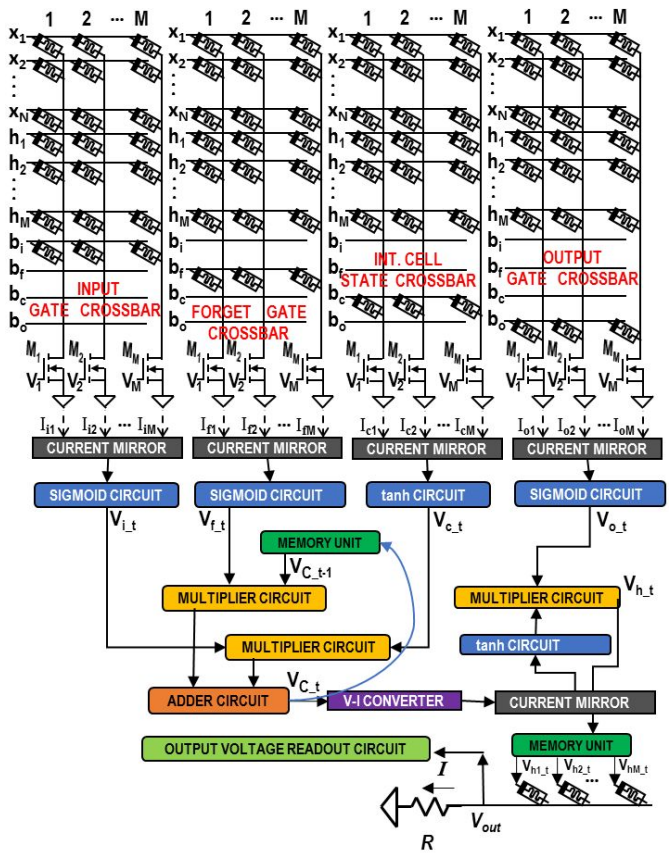

Fig. 3 CMOS-memristive hardware implementation of LSTM cell.